\documentclass[aps,twocolumn,prb,footinbib]{revtex4-1}
\usepackage{graphicx}
\usepackage{lipsum}
\usepackage{amssymb}
\usepackage{amsfonts}
\usepackage{amsbsy}
\usepackage{amsmath}
\usepackage{textcomp}
\usepackage{braket}
\usepackage{bm} 

\DeclareGraphicsExtensions{.pdf,.gif,.jpg}

\begin{document}

\linespread{1.0}

\title{Emergence of Bloch oscillations in one-dimensional systems}

\author{Bogdan Stefan Popescu}
\email[Electronic mail: ]{popescu@udel.edu}
\altaffiliation[\\Present address: ]{Department of Physics and Astronomy, University of Delaware, Newark, DE 19716-2570, USA}

\affiliation{Max Planck Institute for the Physics of Complex Systems, Noethnitzer Str.~38, 01187 Dresden, Germany}

\author{Alexander Croy} 
\email[Electronic mail: ]{alexander.croy@tu-dresden.de}
\affiliation{Institute for Materials Science and Max Bergmann Center of Biomaterials,
Dresden University of Technology,
01062 Dresden, Germany}

\begin{abstract}
Electrons in periodic potentials exhibit oscillatory motion in presence of an electric field. Such
oscillations are known as Bloch oscillations. In this article we theoretically investigate the emergence of
Bloch oscillations for systems where the electric field is confined to a finite region, like in 
typical electronic devices. We use a one-dimensional tight-binding model within the
single-band approximation to numerically study the dynamics of electrons after a sudden switching-on
of the electric field. We find a transition from a regime with direct current to Bloch oscillations when increasing the
system size or decreasing the field strength. We propose a pump-probe scheme to observe the
oscillations by measuring the accumulated charge as a function of the pulse-length.
\end{abstract}

\maketitle

\section{Introduction}
Since the seminal works of Bloch\cite{bloc29} and Zener\cite{zene34} it is well established that 
electrons in a periodic potential exhibit oscillatory motion after a static electric field is
switched on (for a review see Ref.\ \onlinecite{raiz97}). The periodic motion of the electrons in this context is also known as Bloch oscillations (BOs). The period of the oscillations is given by $T_{\rm B} = h/F_0 a$, where $h$ is Planck's constant, $a$ is the lattice spacing and $F_0$ the electric field-strength. In practice
the BOs are typically suppressed by loss of coherence due to collisions, which makes it challenging to observe
BOs in experiments. Over the last decades BOs have been found and characterized in various systems, for instance in semiconductor super-lattices \cite{ohno90a,wasc93a,lyss97a}, ultracold atomic gases \cite{daha96,wilk96a,ande98a} or photonic waveguide arrays \cite{pert99a,mora99a,sapi03a}.

Also on the theoretical side there has been considerable interest in studying BOs. For instance, their interplay with disorder\cite{doma+03,lo10}, light\cite{holt92a,rotv95a}, interactions\cite{buch03a,schu08b,lghi12a} or other degrees of freedom\cite{sisa+16} has been
investigated.
In many cases, the essential physics can already be 
captured using a single-band approximation and considering a one-dimensional (1D) tight-binding (TB) chain\cite{hart04}.

In most theoretical studies BOs have been considered in infinite systems. Here, we investigate the emergence of
BOs in a system where the electric field is confined to a finite-size region. This scenario corresponds to the typical setup of a device coupled to electronic leads. If the device is sufficiently small, one expects to find
a direct current (DC) in the steady state. On the other hand, if the system is infinitely large, one will get BOs as discussed above. In this article we study the transition between the two regimes by increasing the device size and by changing the magnitude of the applied electric field. An important issue is the detection of BOs in such devices. We propose a pump-probe scheme, which uses the rising and falling flanks of a nearly rectangular pulse, to
obtain information about the internal dynamics and BOs.

As in Ref.\ \onlinecite{hart04} we consider a 1D TB system. To account for the time-dependence of the electric field, we use a recently developed numerical method\cite{pope16a}, which is based on the time-dependent non-equilibrium Green's function (TDNEGF) formalism \cite{jauh94,haug96} in combination with an
auxiliary-mode expansion\cite{croy09a}. This approach has already been successfully applied to study time-dependent nanoelectronics\cite{pope12a,chen14c,xie13a,cao15a,wang13a,croy12a,croy12b} and it allows us to obtain
the time-dependent currents and occupations during and after the pulse has been applied. This information
is then used to characterize the spatiotemporal evolution of the electrons in the device.

The article is organized as follows: Section \ref{sec_set_meth} gives details on the specific setup and introduces the numerical scheme used in our study. Section \ref{sec_res_disc} elaborates on the main results, whereas Sec.\ \ref{conclus} concludes the present work.

\section{Transport setup \& Method}\label{sec_set_meth}
\begin{figure}[b]
\includegraphics[width=0.48\textwidth]{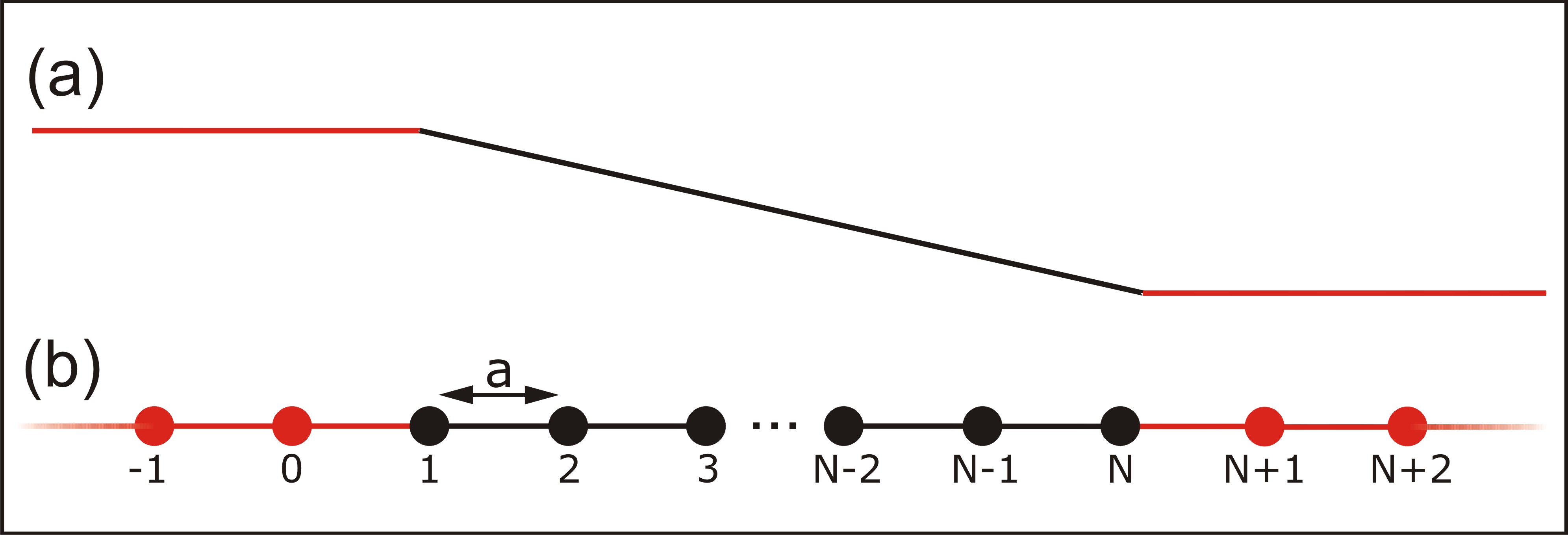}
 \caption{(a) Sketch of the field-induced site-energies (constant in the leads, linearly decreasing in the system). (b) Sketch of the TB chain: the system sites (orbitals) depicted in black (indexed from 1 to $N$) are connected to left and right leads depicted in red.}
 \label{fig1}
\end{figure} 
The setup consists of a generic one-dimensional (1D) system which is described as a tight-binding chain. The system as well as reservoirs sites are coupled to their nearest neighbors (see Fig.\ \ref{fig1}). Thus, the Hamiltonian in second quantization is expressed as 
\begin{equation}
\begin{split}
	&H(t) = \sum_{n=-\infty}^{\infty} \varepsilon_n(t) c^\dagger_n c_n 
				      - \gamma \sum_{n=-\infty}^{\infty} \left( c^\dagger_n c_{n+1} + c^\dagger_{n+1} c_{n} \right) \\
&\mathrm{with}\ \ \ \ \varepsilon_n(t) = \begin{cases}
\varepsilon_L(t)&, \ \ \ \mathrm{if}\ \ n \leq 0 \ \ \mathrm{(left~lead)} \\
\varepsilon_{Dn}(t)&, \ \ \ \mathrm{if}\ \ 1 \leq n \leq N \ \ \mathrm{(device)}\\
\varepsilon_R(t)&, \ \ \ \mathrm{if}\ \   n > N \ \ \mathrm{(right~lead)}~.
\end{cases}
\end{split}
\end{equation}
Here, $\varepsilon_n(t)$ denotes the time-dependent site energies for device and leads, $\gamma$ is the inter-site hopping parameter, whereas $c^\dagger_n$ ($c_n$) creates (annihilates) an electron at site $n$. The device region consists of sites $1, \ldots, N$, while the left (right) lead is defined by sites $n < 1$ ($n > N$).

To study the time-evolution of the device, we employ a numerical method based on the TDNEGF formalism, which was recently put forward in Ref.\ \onlinecite{pope16a}. The time-evolution of the reduced density operator (RDO) ${\bm \sigma}$ is given by
\begin{equation}
  \mathrm{i}\hbar \frac{\mathrm{d}}{\mathrm{d}t} {\bm \sigma}(t) = \big[ {\bf H}_{D} (t) , {\bm \sigma}( t ) \big]     
    +  \mathrm{i} \sum_{\alpha \in \{ {\mathrm{L, R}} \} } \big( {\bf \Pi}_\alpha ( t ) + {\bf \Pi}^\dagger_\alpha ( t ) \big)\;.
    \label{eq:EOMSEDM1}
\end{equation}
Here, ${\bf H}_{D}(t)$ denotes the device part of the Hamiltonian, whereas the current matrices ${\bf \Pi}_\alpha(t)$ are given by another set of equations (for details see Refs.\ \onlinecite{pope16a,croy09a}). These allow it to compute the time-resolved current from reservoir $\alpha=L,R$ into the device via
\begin{equation}
  J_\alpha(t) = \frac{2 e}{\hbar} \mathrm{Re} \mathrm{Tr} \left\{ {\bf \Pi_\alpha} (t) \right\}\;.
\end{equation}  
Note that this propagation scheme solves Eq.~(\ref{eq:EOMSEDM1}) in time-domain. Hence one has access to time-resolved quantities like the instantaneous occupations given by ${\bm \sigma}_{ii}(t),~i=1, \ldots, N$. Moreover, the partial current between next neighboring system sites is given in terms of matrix elements of the RDO by
\begin{equation}
J_{i+1,i}(t) = \mathrm{i} \frac{e \gamma}{h} \big( {\bm \sigma}_{i+1,i}(t) - {\bm \sigma}_{i,i+1}(t) \big)\;.
\end{equation}

\section{Results and Discussion}\label{sec_res_disc}

\subsection{Steady-state dynamics}
\begin{figure}[tb]
 \centering
 \includegraphics[width=0.48\textwidth]{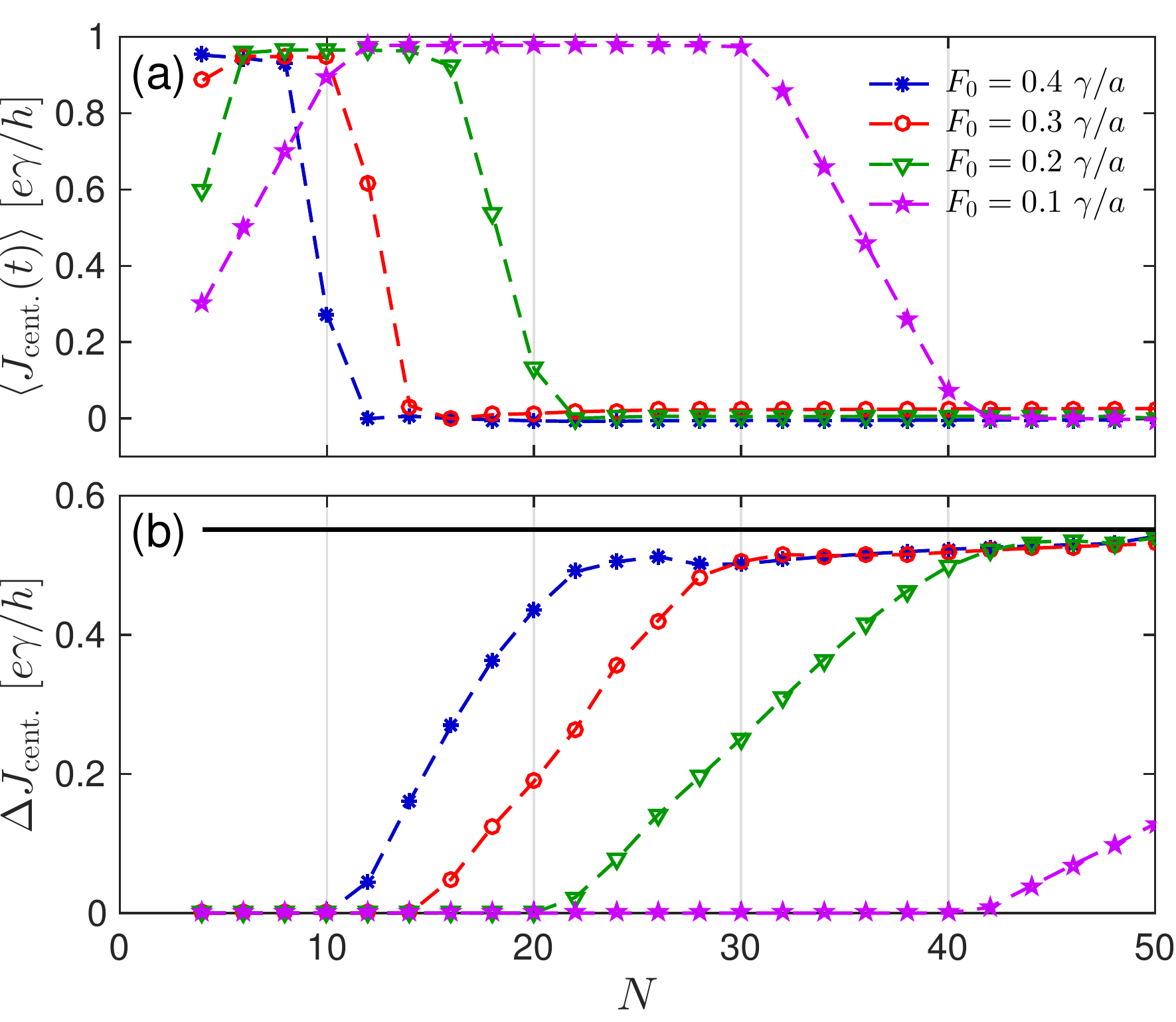}
 \includegraphics[width=0.473\textwidth]{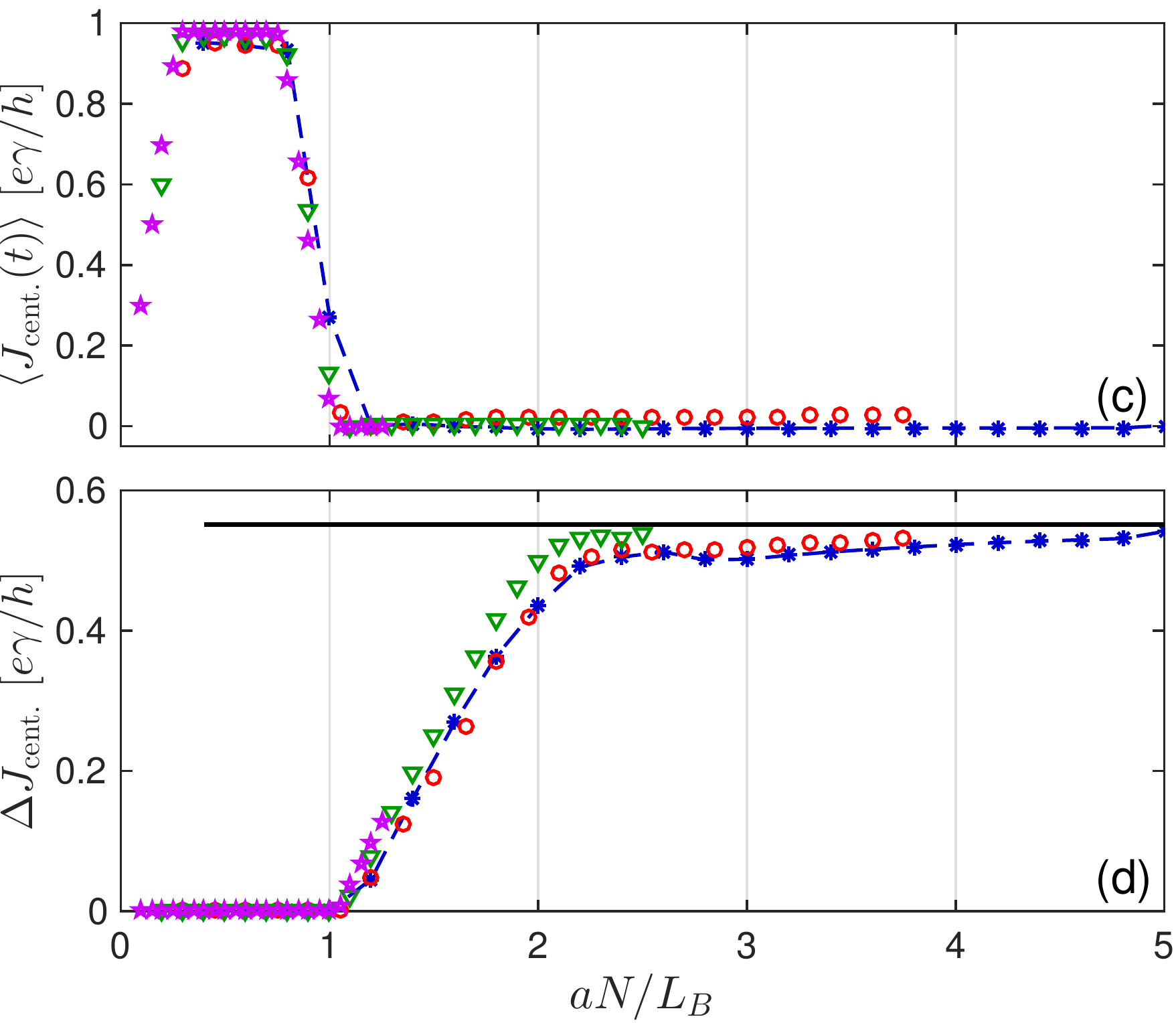}
 \caption{(a) Time-average of the electronic current in the center of the system, $J_{\mathrm{cent.}}$ and (b) half range of mean absolute deviation for $J_{\mathrm{cent.}}$, both computed as  function of system size ($N$) for several values of field magnitude ($F_0$). (c) and (d) show the same data as (a,b) but with the system size being scaled by the Bloch length, $L_B$. Dashed lines are guides to the eye. The solid black lines indicate the current amplitude
 according to Eq.\ \eqref{eq:bo_current}.}
 \label{fig2}
\end{figure}  
In this section we analyze the occurrence of Bloch oscillations in response to the sudden switching-on of an external electric field. Initially, the system is in equilibrium with the reservoirs at $\mu=\mu_{\rm L}=\mu_{\rm R}=-\gamma$ and $k_{\rm B}T=10^{-2}\gamma$. An electric field (or equivalently a bias voltage) is suddenly switched on at $t=t_0$ and it is assumed that the voltage drop is linear across the device. Hence, the time-dependence of the site energies is given by: $\varepsilon_i(t)= a F(t) [(N+1)/2-i]$ and $\varepsilon_{\rm L}=-\varepsilon_{\rm R} = a F(t) (N-1)/2$,
where $F(t) = F_0 [ 1+\text{erf}((t-t_0)/t_{r}) ] / 2$ is the instantaneous force due to an electric field $F(t)/e$ and $\mathrm{erf}$ the error function. The rise time of the electric field is given by $t_{r}$ (in all present simulations $t_{r}=10~\hbar / \gamma$). 

First, in Fig.\ \ref{fig2} we explore the steady-state behavior. To this end we calculate the time-average of the current through the center of the system, $\langle J_{\mathrm{cent.}}(t) \rangle$, where $J_{\mathrm{cent.}}(t)  \equiv J_{N/2+1,N/2}(t)$. As one can see in Fig.\ \ref{fig2}(a), for small system sizes $N$ there is a linear increase of the average current with system size, which eventually saturates to a maximal value. For larger $N$, there is a transition regime, where $\langle J_{\mathrm{cent.}}(t) \rangle$ monotonously decreases. For even larger $N$ the average current becomes identically zero. 
This behavior can readily be understood by noting that the band-width of the leads is $4\gamma$ and the bias-voltage becomes $e V = \varepsilon_{\rm L}-\varepsilon_{\rm R} = F_0 a (N-1)$ for $t-t_0\gg t_r$. The current
increases as $J= (e/h) V$ up to the point where the lower band-edge of the left lead raises over the chemical
potential in the right lead, i.e., for $eV = 2\gamma+\mu$. Now, all occupied states in the left lead have corresponding empty states in the right lead. This changes when the chemical potential in the left lead raises above the upper band-edge in the right lead, i.e., for $eV = 2\gamma-\mu$. The current drops linearly to zero,
$J= (e/h) (4\gamma - eV)$, since at $eV=4\gamma$ the bands of the leads do not overlap any longer.

In addition, in Fig.\ \ref{fig2}(b) we show the half range of the mean absolute deviation for $J_{\mathrm{cent.}}$. This quantity is defined as 
\begin{equation}
\begin{split}
\Delta  J_{\mathrm{cent.}} = \frac{1}{2}\big[ \mathrm{max}&( J_{\mathrm{cent.}}(t) - \langle \mathrm{J}_{\mathrm{cent.}}(t) \rangle) \\ &-\mathrm{min}( \mathrm{J}_{\mathrm{cent.}}(t) - \langle \mathrm{J}_{\mathrm{cent.}}(t) \rangle) \big]~. 
\end{split}
\end{equation}
In the range where the average current is non-vanishing, we find that $\Delta  J_{\mathrm{cent.}}$ is zero. This implies that a DC is flowing in the steady state as long as $F_0 a (N-1) \leq 4\gamma$.
From Fig.\ \ref{fig2}(b) one sees that $\Delta  J_{\mathrm{cent.}} > 0$ if the system size is further increased and
the average current vanishes. This behavior is consistent with an oscillating current.

The transition from DC towards an oscillatory regime can be qualitatively explained by the emergence of BOs. Specifically, there is a typical length scale of BOs, which is given by the Bloch length\cite{hart04}
\begin{equation}
    L_{\rm B}  =  \frac{4\gamma}{F_0}\;,
\end{equation}
where $4\gamma$ is the band width of a 1D-TB chain. This means that the system size needs be greater than the Bloch length, in order to observe oscillatory behavior. This argument is supported by Figs.\ \ref{fig2}(c-d), where the scaling of the data by the Bloch length is shown. 

\begin{figure}[]
 \centering
 \includegraphics[width=0.489\textwidth]{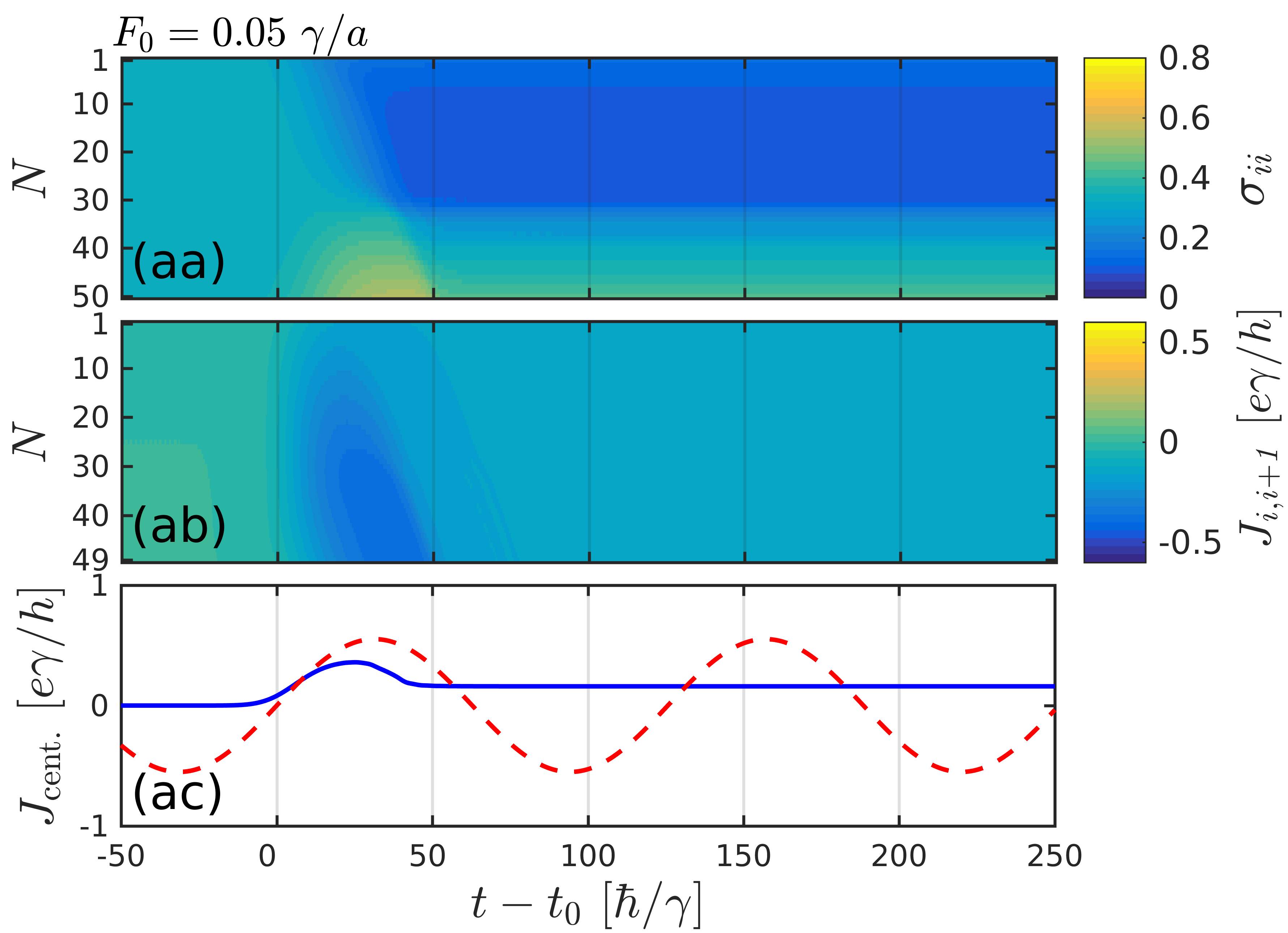}
  \includegraphics[width=0.489\textwidth]{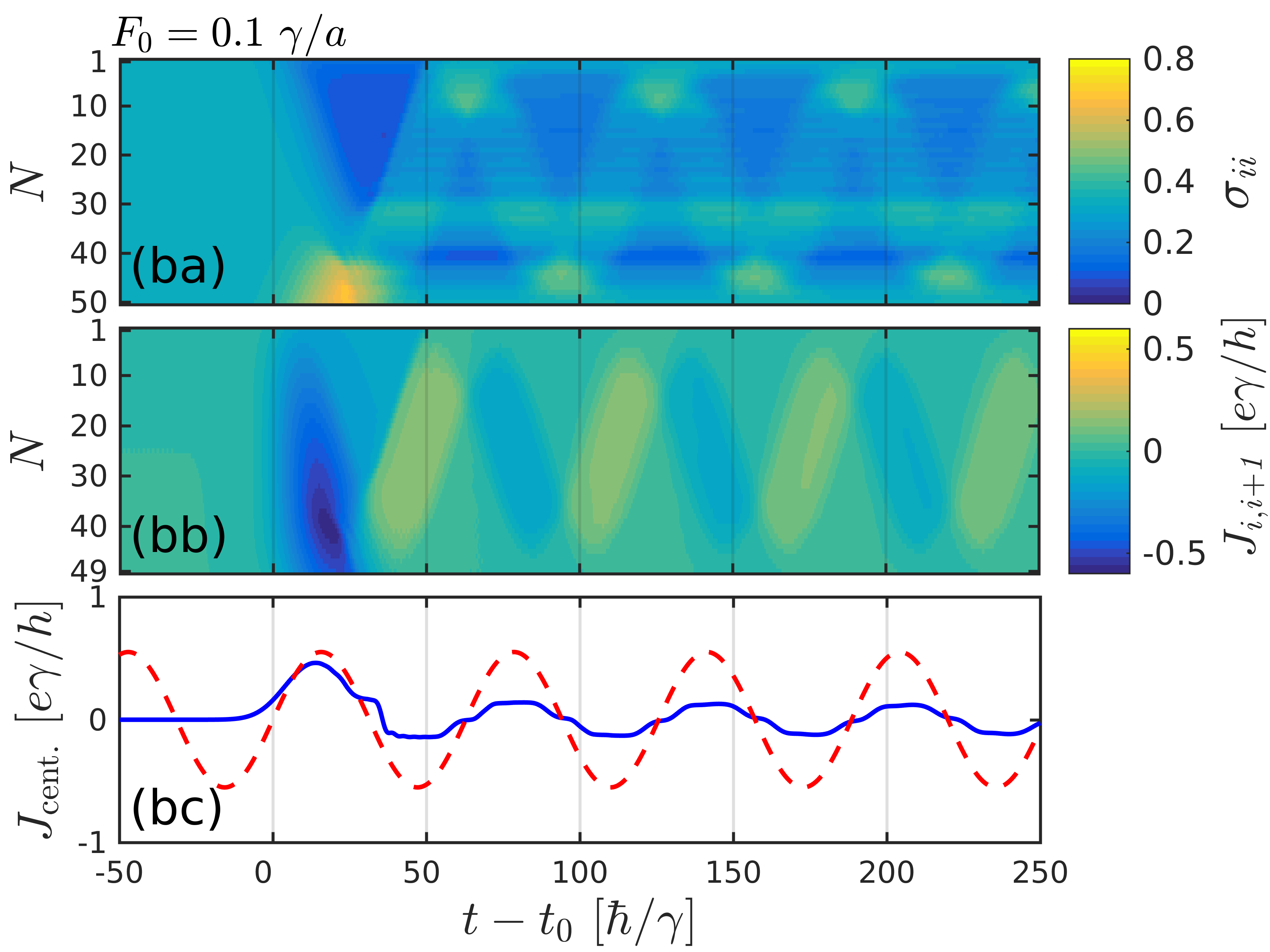}
   \includegraphics[width=0.489\textwidth]{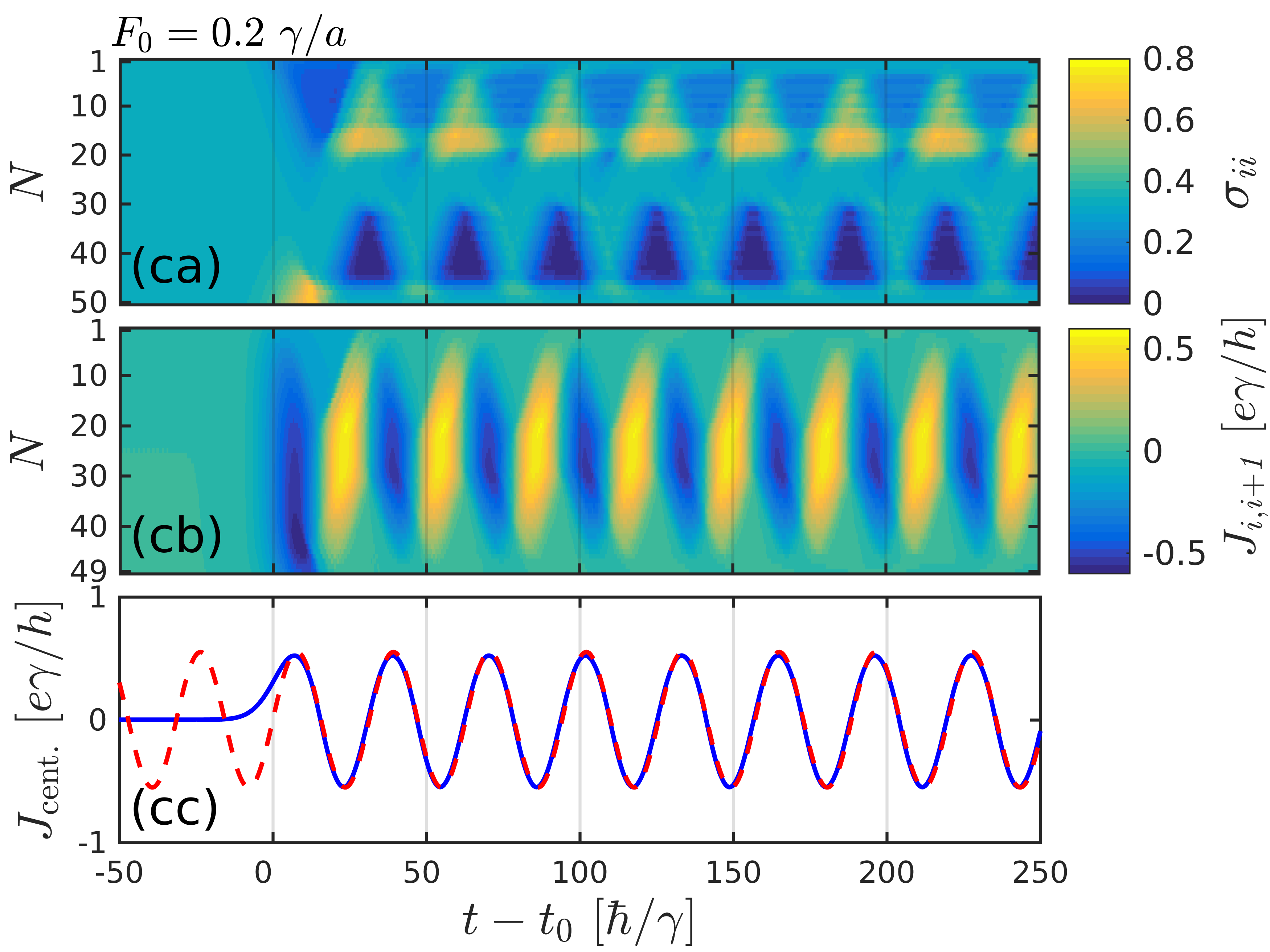} 
   \caption{Conceptually different transport regimes obtained by changing $F_0$. (aa-ac)  $F_0=0.05\gamma/a$: Typical transient-relaxation to DC behavior. (ba-bc) $F_0=0.1\gamma/a$: Intermediate regime with onset of BOs. (ca-cc) $F_0=0.2\gamma/a$: Regime of BOs. For each of the triple blocks, the upper plot shows the occupation number, the middle one depicts the current between neighboring sites and the lower plot displays the current in the center of the system, $J_{\mathrm{cent.}}(t)$. The system size is $N=50$ in each case. The (red) dashed lines
   show the current according to Eq.\ \eqref{eq:bo_current}.}
 \label{fig3}
\end{figure}
\subsection{Time-resolved dynamics}
To explore the effect in more detail, we fix the system size to $N=50$ and examine the time-dependent response for different external field magnitudes. Specifically, the time-dependence of the site occupation numbers and the currents obtained by our method are shown in Fig.\ \ref{fig3}. Each triple-block plot shows the dynamics computed for a value of $F_0$, namely $F_0=0.05\gamma/a,~ 0.1\gamma/a$ and $0.2\gamma/a$. Note that in each block displayed are from top to bottom: the site occupation number, $\sigma_{ii}(t)$ (given by the diagonal elements of the RDO), the partial currents between neighboring sites, $J_{i,i+1}(t)$, as well as the current through the center of the system, $J_{\mathrm{cent.}}(t)$. 

In the range of small $F_0$ values, the electron transport through the device is characterized by transient regime towards a steady state with a DC and constant occupations as explained above. This is shown in Fig.\ \ref{fig3}(aa-ac). For intermediate field strengths the onset of oscillations can be observed in both, the population dynamics, i.e., Fig.\ \ref{fig3}(ba), as well as in the time-dependence of the partial currents, displayed in Fig.\ \ref{fig3}(bb,bc). Lastly, the lower-most triple block depicts data obtained for $F_0=0.2\gamma/a$. Here the oscillatory behavior acquires perfect sinusoidal form, as can be seen in Fig.\ \ref{fig3}(cb,cc). Comparing with the expression for the current obtained by a semi-classical analysis (see Appendix \ref{sec:semi}), we find an excellent agreement.

\begin{figure}[tb]
 \centering
 \includegraphics[width=0.48\textwidth]{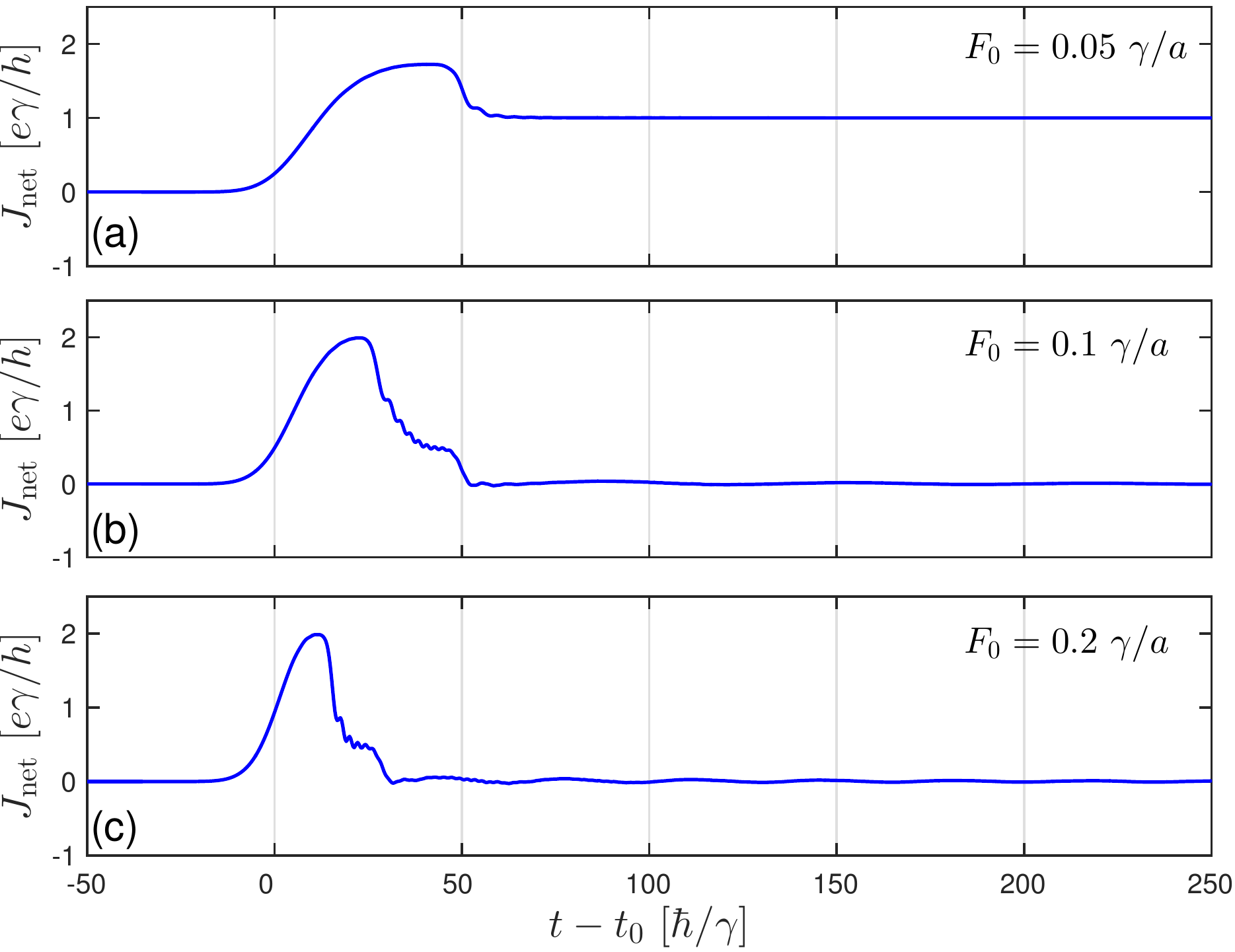}
 \caption{Net current through the system ($N=50$) for the three field magnitudes used for Fig.\ \ref{fig3}. (a) shows the transient transition towards a stationary state with non-vanishing DC current, while (b) and (c) correspond to the regime of BOs. In the latter regime, the net current vanishes.}
 \label{fig4}
\end{figure} 
The transition from DC to BOs regime can also be seen in the net current through the device, $J_\mathrm{net} \equiv (J_L-J_R)/2$. The latter is shown in Fig.\ \ref{fig4}, computed for each of the above field magnitudes. Namely, the upper plot in Fig.\ \ref{fig4}(a) captures the DC regime characterized by a non-vanishing $J_{\mathrm{net}}$. In contrast, Figs.\ \ref{fig4}(b,c) show the emergence of BOs. Here, after the transient response to the switching of the field, the net current vanishes eventually. This situation corresponds to the occurrence of BOs seen in Fig.\ \ref{fig3}(cc).

\subsection{Transferred charge}
\begin{figure}[tb]
 \centering
 \includegraphics[width=0.49\textwidth]{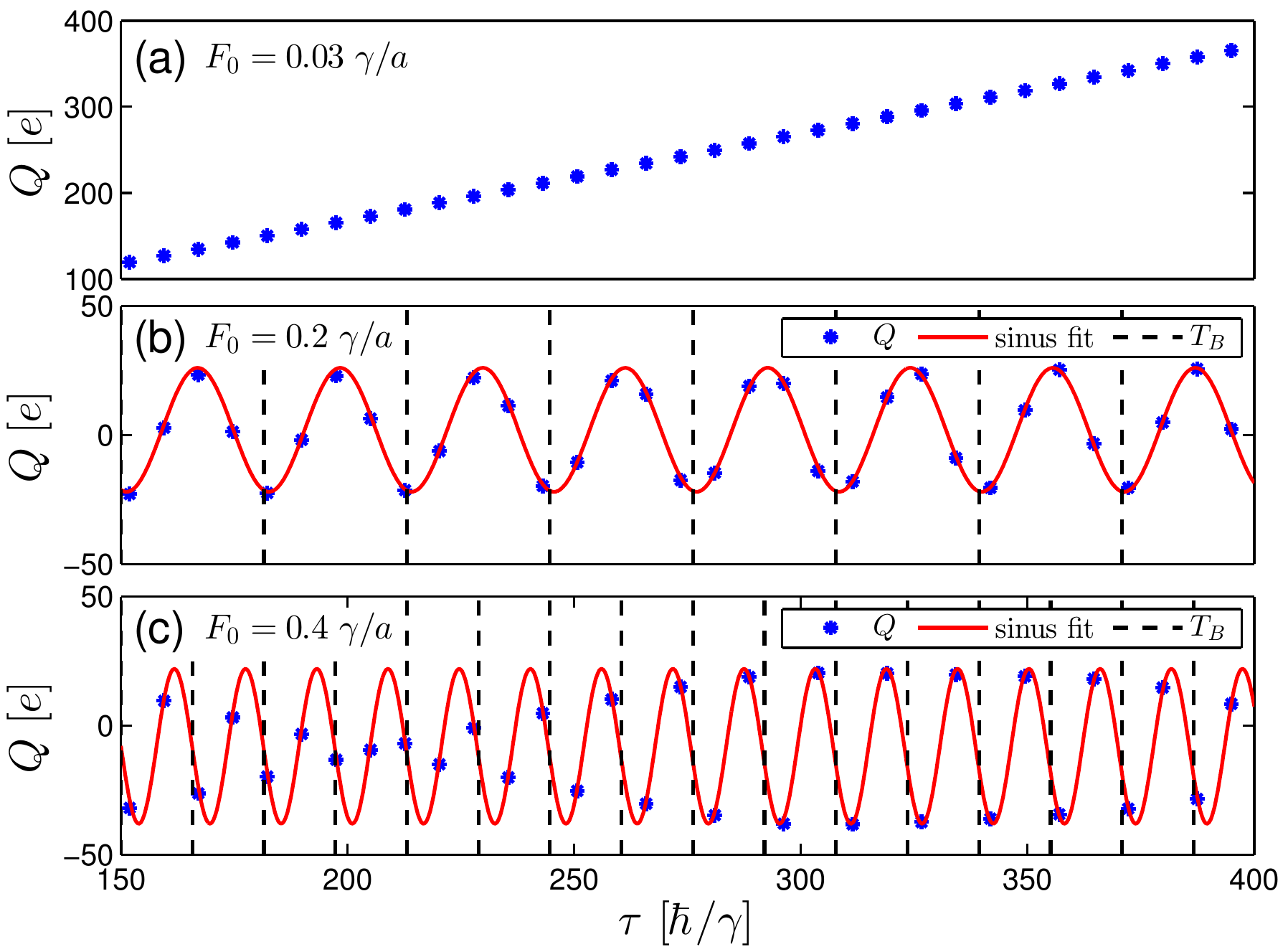}
 \caption{Accumulated charge $Q$ detected at the right electrode as function of pulse-length $\tau$ for three values of field strength $F_0$ corresponding to (a) direct current behavior and (b,c) BOs. For (b) and (c) the data points are fitted by a sine function (red curve). The vertical dashed lines mark the period of BOs.}
 \label{fig5}
\end{figure} 
In an experimental setting the internal currents or the instantaneous occupation number are typically not directly measurable. While the data shown in the previous section clearly indicate the presence of two distinct transport regimes within the quantum system, it is desirable to visualize the pattern of BOs also in experimentally-accessible observables. As we show in the following, a pump-probe scenario, where the rising flank of the pulse is used as a pump and the falling flank as a probe, can be utilized to investigate the internal dynamics. This approach has been, for example, successfully used for studying the dynamics of charge qubits in double quantum-dots\cite{hafu+03,croy11a}.
Accordingly, the time-dependence of the electric field is explicitly given by
$F(t) = F_0 [ \text{erf}((t-t_0)/t_{r}) - \text{erf}((t-t_0-\tau)/t_{r})  ] / 2$, where $\tau$ is the pulse
length. We calculate the charge injected into the right electrode, which is given by $Q(\tau) = -\int_{-\infty}^{\infty} \mathrm{d}t' J_{\rm R}(t')$, as a function of the pulse-length\cite{foot1}. 

In the DC regime, one expects
$Q(\tau)$ to increase linearly with $\tau$. This is confirmed by Fig.\ \ref{fig5}(a), which shows $Q(\tau)$ for the field-strength $F_0=0.03\gamma/a$. The transient response seen in Fig.\ \ref{fig4}(a) is only relevant for small $\tau<t_{\rm trans}\approx 60 \hbar/\gamma$, which is not shown here.
On the other hand, in the BOs regime $Q(\tau)$ is entirely determined by the transient responses at the instants of switching on and off of the field. The switching-on contribution is the same for all $\tau>t_{\rm trans}$ and leads to a constant offset. The switching-off contribution depends on the state of the system\cite{croy11a} and will thus be modulated by the BOs. This can be seen in Figs.\ \ref{fig5}(b,c) which show the pulse-length dependence of $Q$ for $F_0=0.2\gamma/a$ and $F_0=0.4\gamma/a$. The accumulated charge can be well fitted by a sine function which has the period $T_B$ of the BOs. The amplitude of the oscillations
depends on the state of the system around the instant of switching-off the pulse.

\section{Conclusions}\label{conclus}
In summary, we studied the emergence of BOs in paradigmatic 1D systems after switching on an external electric field. For small system sizes, $N\ll L_{\rm B}/a$, no BOs are observed and the stationary current is directed as expected.
Increasing the system size beyond the Bloch-length $L_{\rm B}$ leads to the emergence of BOs, which are characterized by oscillating internal currents and a vanishing net current through the device.
We find that the system size $a N$ has to be larger than $2 L_{\rm B}$ to obtain BOs corresponding to
the ones found in a homogeneous field.
By employing a pump-probe scheme, which uses the rising and falling flanks of a nearly-rectangular pulse, we showed
that the BOs can be identified in the accumulated charge obtained as a function of pulse-length. 
This offers a route to study dynamics of BOs and their interplay with noise and/or other degrees of freedom in nano-scale devices.\\

\vspace{-0.7cm}

\appendix 
\section{Semi-classical description of BOs}\label{sec:semi}
For an infinite system one can describe the electron dynamics using a semiclassical approach\cite{ashc76a,hart04}. An electron is characterized by
its position $x(t)$ and wave-vector $k(t)$, which obey the equations of motion 
\begin{equation}
	\frac{\partial}{\partial t}x(t) = \frac{1}{\hbar} \frac{\partial \epsilon(k(t)) }{\partial k}, ~~~ \frac{\partial}{\partial t}k(t) = -\frac{1}{\hbar}F(t)~.
\end{equation}
The band-structure for the 1D tight-binding chain is $\epsilon(k) = -2\gamma \cos(k a)$. Before the electric field is switched on, the state $\ket{k_0}$ is occupied with probability
$f(\epsilon(k_0))$, where $f$ is the Fermi function. After an instantaneous switching of the (homogeneous) electric field at time $t_0=0$ from zero to $F_0$, one finds for the time-dependence of $x$ and $k$,
\begin{align}\label{bo:sol}
    x(t) ={}& x_0 - \frac{2\gamma}{F_0} \cos(k_0 a) + \frac{2\gamma}{F_0} \cos[k_0 a - \omega_{\rm B} t]\;,\\
    k(t) ={}& k_0 - \frac{\omega_{\rm B}}{a} t\;,
\end{align}
where $\omega_{\rm B} = 2\pi/T_{\rm B}$ is the Bloch-frequency.
Consequently, the current is given by
\begin{align}
    J(t) ={}& e \int^{\pi/a}_{-\pi/a} \frac{dk_0}{2\pi} f(\epsilon(k_0)) \left.\frac{\partial}{\partial t}x(t) \right|_{k(t)=k_0 - \omega_{\rm B} t/a} \notag\\
    ={}& -\frac{4 e \gamma}{2\pi \hbar} \sin(k_{\rm F} a)\sin({\omega_{\rm B} t})\;.\label{eq:bo_current}
\end{align}
In the last step above zero temperature was assumed and the Fermi wave-vector $k_\mathrm{ F}$ is related to the
Fermi energy via $\epsilon_\mathrm{F} = \epsilon(k_\mathrm{ F})$.
 


\providecommand{\newblock}{}

\bibliographystyle{iopart-num}
\end{document}